# The Effective Cross-sections of a Lensing-galaxy: Singular Isothermal Sphere with External Shear

*Dong-Wook Lee\* and Sang-Joon Kim*
[ *\*E-mail: dr.dwlee at gmail.com* ]

Dept. of Astronomy & Space Science, Kyung-Hee University,
Seochon-dong 1, Yongin, Gyeonggi-do, 446-701,
*Republic of Korea.*

### - ABSTRACT -

Numerical studies on the imaging and caustic properties of the singular isothermal sphere (*SIS*) under a wide range of external shear (from 0.0 to 2.0) are presented. Using a direct inverse-mapping formula for this lens system (Lee 2003), we investigate various lensing properties under both a low (i.e., $\gamma < 1.0$), and a high (i.e., $\gamma > 1.0$) shear case: image separations, total or individual magnifications, flux ratios of 2-images, maximum number of images, and lensing cross-sections. We systematically analyze *the effective lensing* cross-sections of *double-lensing* and *quad-lensing* systems based on the radio luminosity function obtained by *Jodrell-VLA Astrometric Survey* (*JVAS*) and *Cosmic Lens ALL-Sky Survey* (*CLASS*). We find that the limit of a survey selection bias (i.e., between a brighter- and a fainter-image) preferentially reduces *the effective lensing* cross-sections of 2-image lensing systems. By considering the effects of survey selection bias, we demonstrate that the long standing anomaly *on the high Quads-to-Doubles ratios (i.e., JVAS & CLASS*: 50% ~ 70%) can be explained by the moderate effective shear of 0.16 ~ 0.18, which is half of previous estimates. The derived inverse mapping formula could facilitate the *SIS + shear* lens model to be useful for galaxy-lensing simulations.

**\*Key-words:** *Cosmology; Theory; Gravitational lens: Galaxies, Cross-section, Magnification bias, Inversion.*

# 1. Introduction

The gravitational *macro-lensing* of a galaxy is distinguished from the *micro-lensing* of a star due to its extended nature of lensing object and its apparent large angular scales of multiple images. The galaxy-lens models can be classified into two groups in general: singular isothermal sphere (e.g. Schneider et al.1993; *hereafter, SIS*) lens models that have a projected circular symmetry and elliptical potential (or mass) lens models (e.g. Blandford & Kochanek 1987; Schramm 1990; Kormann et al. 1993; Kassiola & Kovner 1993; *hereafter, SIE*). Compared to elliptical lens models, the *SIS*-model has a merit of simplicity of mathematics due to its radial symmetry. Using this representative galaxy-lens model, the *SIS + shear* model, we extensively investigate in depth their properties related to imaging, magnification, cross-sections, caustics, and multiple images.

Extra-galaxies are considered to have an extensive dark halo, regardless of their light emitting morphology. Considering the presence of large dark halos in lensing galaxies, the assumption of radial symmetry for a lensing galaxy is generally valid at first order as used by many cosmological lensing statistics and N-body galaxy-lens studies (Turner et al. 1984; Katz & Paczynski 1985; Claeskens et al. 2001; Mitchell et al. 2005). Even from many individual galaxy-lens modeling studies, the *SIS* property for deflection angles has been confirmed up to ~10 Kpc from the galactic center in other studies (e.g. Koopmans et al. 2009). The isothermal property for lensing galaxies has also been widely accepted (e.g. Antonio et al. 2011) based on *the SDSS survey* (Oguri et al. 2006 & 2008) and *the Sloan Lens ACS survey* (Bolton et al. 2006).

Although the *SIS*-lens model is originally an ideal galaxy-lens, it provides most of the general features for galaxy lensing. When modeling an individual quasar-galaxy lensing system with observational constraints, it is necessary to include more parameters in order to get a finer model within observational errors. The pure *SIS*-model is inappropriate for the fine tuning of individual lens-systems since it does not account for the 3-image nor 4-image lens systems. It is thus necessary to add an external tidal shear to the pure *SIS*-lens model (Kovner 1987) so as to get fine modeling of individual QSO-lens systems.

On the other hand, this kind of external perturbation can be similarly treated by considering an internal distortion of the lensing potential itself (i.e., ellipticity of a lensing galaxy). Thus, elliptical models for lensing potential (Blandford & Kochanek 1987; Schramm 1990) and for mass distributions have been developed by others (Kormann et al. 1993; Kassiola & Kovner 1993). Although the elliptical models have quite similar properties of caustics like the *SIS+shear* models, the elliptical lens models, however, have more complexities of mathematical calculations in obtaining the deflection angle and lensed images.

By adding the external shear effect on a pure *SIS*-lens system, the *SIS + shear* model becomes comparable to the *SIE* model in terms of the number of parameters and the properties of caustics and multiple images (e.g. Keeton, Kochanek & Seljack 1997). It is well known that both the *SIS + shear* model and the *SIE* models have produced very similar fits for many individual galaxy-lens modeling (*refer to CASTLEs*[1]). However, more complex models are usually required for fine model fitting. In the source plane, it is known that the caustic properties of the *SIS + shear* model are known to have almost the same features of the *SIE* model while keeping the relation between the shear and the ellipticity (i.e., $\gamma \sim \epsilon/3$; Keeton, Kochanek, & Seljack 1997). Even if their lens-plane characters (i.e., potential and mass) are somewhat different in reality (Kovner 1987; Schramm 1990), the *SIS + shear* model can produce a double lensing, a marginal lensing (i.e., 3-images) in a naked-cusp caustic and a quad-lensing (i.e., 4-images) in a tangential diamond caustic, just like other elliptical *SIE* models.

In dealing with N-body galaxy-lens simulations or cosmological lensing statistics, one needs to use a systematically coherent base model for the lensing galaxy. In such a case, the pure *SIS* model without the shear has been generally used over 3 decades due to its simplicity with (or without) its radial truncation at large radii (e.g. Turner et al. 1984; Katz & Paczynski 1985; Claeskens et al. 2001; Mitchell et al. 2005; Wong et al. 2011). For these kinds of N-body galaxy-lens simulations, however, adding an external shear to the pure *SIS* model can be the simplest way to improve the realities of N-body galaxy-lensing. Since the additional shear can easily reproduce almost similar effects like the role of ellipticity in galaxy lensing, the shear can be easily translated into the ellipticity between the *SIE* model and the *SIS+shear* model.

Compared to the elliptical lens models, however, the *SIS + shear* model is more convenient in dealing with heavy numerical simulations due to its own radial symmetry, even though their lensing properties are quite similar. In galaxy lensing problems, the caustic is an important issue for both the cross-section problem and the magnification related problem. These issues have been widely studied by others so far. For example, Keeton, Gaudi, & Petters (2003) presented the flux anomaly study for cusp caustic lensing and suggested sub-structure for lensing potential. Keeton, Gaudi, & Petters (2005) studied the flux anomaly in fold lenses based on the fold relation. Rozo, Chen & Zentner (2013) considered the possible effects of tri-axiality of lensing halos for the *Quads-to-Doubles ratio* and showed that tri-axiality can affect the lensing cross-sections.

---

[1] The CASTLEs project (http://www.cfa.harvard.edu/castles)

In gravitational lens surveys there has been a long standing problem of the excess frequency of 4-image systems compared to 2-image systems. This is the so called, *high Quads-to-Doubles ratio* (e.g. King & Browne 1996; Rusin & Tegmark 2001). This anomaly has been evidently observed especially by two independent radio surveys, also known as the *JVAS & CLASS* radio lens surveys (Patnaik et al 1992; King et al. 1999; Phillips et al. 2001; Browne et al. 2003) of which those frequencies between 4-image systems and 2-image systems are more than 50%, or up to 70%, respectively. Also, optical lens surveys produced a similar trend for the *Quads-to-Doubles ratio* which is about 30% (*refer to CASTLEs*), though the degree of anomaly is not as strong as that of radio lens surveys.

The observed *high Quads-to-Doubles ratios* in radio lens surveys have puzzled astronomers over a decade since its degree is almost double that of theoretical expectation (King & Browne 1996; Rusin & Tegmark 2001; Keeton et al. 2001). In order to cope with this anomaly, several attempts have been proposed; for example, by considering a possible additional shear effect of line-of-sight (*hereafter, LOS*) galaxies (Holder & Schechter 2003; Dalar & Watson 2004; Wong et al. 2011), of neighboring satellite galaxies (Cohn & Kochanek 2004), and of a neighboring cluster of galaxies around a main lensing galaxy (Keeton & Zabuldoff 2004). All these approaches took into account additional shear contributions to a main lensing galaxy in order to increase the lensing cross-section.

Another possibility might be that statistical sampling is not enough to draw a significant conclusion yet. Up to now, however, strong lensing systems for quasars are more than 100 (refer to the list of *CASTLEs*) of which observational possibility is quite rare in the order of 1/1,000 (Turner et al. 1984). Thus, it would mean that finding out more than 100 multiply lensed quasar systems can already tell us something as many authors have tried to make sense out of them. Otherwise, as Rusin & Tegmark (2001) mentioned; *"Could there exist some observational bias which favors the discovery of 4-image systems over 2-image systems?".* In this work, we investigate the same question in terms of model consideration which can suppress the observational frequency of 2-image lensing systems.

In this work we simply regard the lensing galaxy itself as a dark halo model, regardless of its detail morphology. We have thus solely based our work on the representative *SIS + shear* model. Based on this representative model, we calculated the image-/caustic properties and analyzed its lensing cross-sections in the source plane, as the cross-section is the fundamental term of lensing statistics. Although the *SIS + shear* model has often been used, it is a little surprising that except for a few theoretical studies (e.g. Kovner 1987), there has been no detailed study on its imaging/caustic properties and cross-section variations for a wide range of external shear values.

We investigate its various lensing properties with a wide range of external shear being from 0 to 2.0. For the cases of cross-section analysis, Finch et al. (2002) analyzed the variation of cross-section of *SIS + shear* model only for low-shear cases (0~1) and derived some analytic expressions for mean magnification and cross-section, which are valid under the maximum shear limit of 1/3. To our knowledge, especially for high shear conditions (i.e., shear > 1.0), any properties of multiple-images and caustics of galaxy-lens model have not been reported in previous studies, except for some specific cases of point-mass lens: *Chang-Refsdal* lens (Chang & Refsdal 1979; Chang 1981) and the binary lens plus shear model (Grieger et al. 1989; Lee 1997; Lee et al. 1998).

Adopting Monte-Carlo schemes with the inverse mapping formula (*see* Section 2) of the *SIS + shear* model, we carried out extensive numerical investigations for various imaging properties including cross-section variations under a wide range of external shears from 0.0 to 2.0. Because this inversion formula is valid for any caustic configuration with shear values, we introduced their imaging properties not only for a low-shear case (i.e., shear < 1.0), but also for a high-shear case (i.e., shear > 1.0) for the first time.

In Section 2, we describe the lens equation and its inverse mapping formula, numerical techniques and Monte-Carlo results for the imaging properties of *SIS + shear* lens systems. One may note that we present these results by comparing the differences between low-shear cases and high-shear cases throughout this work.

In Section 3, multiple images for an extended source are presented with their caustic changes depending on the external shear, including the high-shear conditions.

In Section 4, *the differential-* and *the effective lensing* cross-sections are presented for various shear values. Depending on the shear parameter and the magnification bias with a survey selection limit, we calculate their cross-section variations and compare their properties. We present the finally estimated observational frequencies of *Quads-to-Doubles* and *Triples-to-Doubles* as a function of the external shear based on the analysis of *the effective lensing cross-section*.

In Section 5, we summarize our work and present a discussion on previous studies along with our new results and suggest possible applications of our work to future N-body galaxy-lensing simulations.

## 2. Lens Equation and Inverse Mapping

### 2.1 Overview

Although there are many gravitational lens models for both point-mass models (i.e., micro-lensing) and extended-lens models (i.e., galaxy-lensing), only a few inverse mapping formulas have been derived so far, such as a point-mass lens, which has a maximum 2 images (Refsdal 1964), the *Chang-Refsdal* lens, which has a maximum 4 images (Chang 1981), a binary point mass lens, which has a maximum 5 images (Schneider & Weiss 1993), and a *SIS*-model without an external shear, which has a maximum 2 images (Schneider et al. 1993).

To our knowledge, these are all the lens models that their analytic inversion formulas have been derived, which can generate the direct inversion from a point in the source-plane to multiple-points in the lens-plane. Additionally, a direct inverse mapping formula was also derived in the case of a binary point mass plus shear model with an extended source (Lee 1997; Lee, Kim & Chang 1998).

As for the other complex-lens models, obtaining the multiple images usually requires time consuming numerical techniques, such as ray-shooting methods (Young 1981; Kayser et al. 1986) or pixelated grid searching techniques (e.g. Keeton 2001) based on the optical reverse-ability in gravitational lensing. However, these indirect numerical techniques cannot provide the exact inversion mapping from a source-point to image-points, and are usually not effective for the problems with an extended source.

Thus, finding these exact points of the multiple-image boundaries with high accuracy is practically impossible when we use these fully numerical methods. Thus, the derivation of a direct inverse mapping formula is an essential way to produce the exact boundary points of lensed images and to straightforwardly calculate the cross-section without consuming significant computing time.

### 2.2 Lens Equation

In order to evaluate the lens equation of the *SIS + shear* model, we arrange the shear orientation angle to the Y-axis following the conventional North-East system as shown in equation 1. The general lens equation of the *SIS + shear* lens system can then be described in terms of a length scale as follows;

$$\bar{S} = \begin{pmatrix} 1 & -\Gamma_1 \\ 1 & -\Gamma_2 \end{pmatrix} \bar{X} - \varphi_{\text{lens}} \left( \frac{\bar{X}}{|\bar{X}|} \right). \tag{1}$$

where, $\Gamma_1 = (1-\kappa-\gamma)$ and $\Gamma_2 = (1-\kappa+\gamma)$ are extra-focusing terms of the external convergence ($\kappa$) and the external shear ($\gamma$), $\bar{S}(S_x, S_y)$, and $\bar{X}(X, Y)$ are 2D vectors in the source- and the lens-plane, respectively, $\varphi$ is the Einstein radius unit of a lensing potential ($\varphi_{\text{lens}} = 4\pi\sigma^2/c^2 \, (D_{LS} \, D_L/D_S)$), $\sigma$ is the line of sight velocity dispersion of a lensing galaxy, and $D_{LS}$, $D_L$, and $D_S$ are the angular diameter distances in cosmology; between the lens and the source, to the lens, and to the source, respectively. Note that we neglect the external convergence effect in this work and the details of the cosmological model do not affect our results.

### 2.3 Inverse Mapping Formula

Another important merit of the *SIS + shear* model compared to the other *SIE* models is that there is a direct inverse mapping formula which enables one to directly derive multiple-point images corresponding to a point source. Since there is no such a direct inverse mapping solution for the cases of *SIE* models, it should therefore be the most useful merit of the *SIS + shear* model among the galaxy-lens models.

If the aim of research is not exact model fitting, but focusing on a global statistical purpose, then one can make use of the *SIS + shear* model as a representative lens model instead of using a pure *SIS* model, which has been widely used so far (Turner et al. 1984; Katz & Paczynski 1985; Claeskens et al. 2001; Mitchell et al. 2005; Wong et al. 2011). Therefore, it seems useful for us to investigate the imaging and caustic properties of this *SIS + shear* lens model in depth with the direct inverse mapping formula.

The direct inverse mapping formula (I.M.F; Lee 2003) for the *SIS + shear* model and its coefficients are presented in equation 2, which was originally derived in the form of $4^{th}$ order polynomials after considerable arithmetic manipulations. Such an inversion technique was originally derived in the *Chang-Refsdal* lens model (Chang & Refsdal 1979; Chang 1981). The formula in equation 2 is for a general inversion solution, so that it can directly solve for multiple-image positions from any given point-source in the source-plane under any external shear condition using a simple root finding method. This is a great merit compared to other numerical grid searching methods (Lee 2003). Once the image position is found, possible imaginary solutions are then checked out with a predefined numerical accuracy.

Here, we present the inverse mapping formula of the *SIS + shear* model and its coefficients (Lee 2003) as follows;

$$\mathrm{I.M.F(X)} \equiv C_1 \cdot X^4 + C_2 \cdot X^3 + C_3 \cdot X^2 + C_4 \cdot X + C_5,$$

where,
$C_1 = (2\gamma(1-\gamma))^2,$
$C_2 = (4\gamma S_x(1-\gamma)^2 - 8\gamma^2(1-\gamma)S_x),$  (2)
$C_3 = ((1-\gamma)^2(S_x^2 + S_y^2) - 8\gamma(1-\gamma)S_x^2 + 4(\gamma S_x)^2 - (2\gamma\varphi)^2),$
$C_4 = (-2(1-\gamma)(S_x^2 + S_y^2) + 4\gamma S_x^2 - 4\gamma\varphi^2)S_x,$
$C_5 = (S_x^2(S_x^2 + S_y^2 - \varphi^2)).$

The corresponding additional pair position of an each image can be obtained by the following equation,

$$Y_i = \frac{S_y \cdot X_i}{(2\gamma \cdot X_i + S_x)}.$$  (3)

By solving this 4$^{th}$ order polynomial equation, one can obtain up to 4 multiple pairs of image positions (i.e., $X_i$ and $Y_i$) in the *SIS + shear* model. Notice that since this inversion equation intentionally includes the lensing potential term, this formula can be directly used even for N-body lensing galaxy problems with different amplitudes of lensing potentials. In such N-body lensing cases, one can even solve the multiple image positions of an individual main lensing galaxy after calculating the external shear amplitude from its neighboring galaxies, compared to the previous case of using the ray-shooting method in N-body galaxy-lensing simulations (e.g. Claeskens et al. 2001).

The magnification (μ) of a lensed image is then calculated by using a Jacobian transformation of the lens mapping (Refsdal 1964; Chang & Refsdal 1979) and its image parity is also determined. The critical curve formula in the lens plane is straightforwardly obtained with equation 4;

$$\mu = \det\left(\frac{\partial \overline{S}}{\partial \overline{X}}\right)^{-1}.$$  (4)

Depending on the image multiplicity and the shear, total/individual magnification probability distributions and image separations are calculated by Monte-Carlo simulations. When calculating their magnification probability distributions, we set a maximum limit of 100, which is required for numerical computation only. Using 10-million random positions in the source plane, we derive its integral cross-section. Our Monte-Carlo simulations resulted in a fractional error of 0.12% for the worst case.

Note that we do not consider any further cosmological dependency. Since our aim was focused on precise calculations of the lensing cross-section with the shear, which is the fundamental element in gravitational lensing statistics, we made use of the normalized lensing properties of a singular isothermal lensing galaxy depending on the external shear parameter. We should notice that this shear term can always be translated into the ellipticity in lensing problems based on *the shear and ellipticity relation* (i.e., $\gamma \sim \epsilon/3$; Keeton, Kochanek, & Seljack 1997).

## 2.4 Imaging Properties

The imaging properties of 2- and 4-image lensing (or 3-image lensing for high shear) events were calculated as a function of the shear parameter by Monte-Carlo methods and their results are presented in *Figure* 1. The magnification probability distributions for different shear parameters are shown in the upper panel and the probability distributions of image separation are displayed in the lower panel. We defined the image separation as the largest observable separation of multiple images as slimily used by Huterer et al. (2004).

The differences of imaging characteristics for high- and low-shear cases can be understood by comparing the two groups in *Figure* 1 (i.e., low-shear (< 1.0); *right-side* vs. high-shear (> 1.0); *left-side*). The differences in their magnification probability distributions are noticeable between a low-shear and a high-shear case, as presented in the upper panel in *Figure* 1. Compared to the 2-image lensing systems of the zero shear, of the low-shear, and of the high-shear cases show different behaviors, in which a high-shear case produces a relatively lower total magnification. When the shear is in low magnitude, the total magnification of 2-image lensing tends to be gradually decreased as the shear increases. However, when the external shear is increased above 1.0 in high magnitude, this trend is reversed.

The maximum separation of 2-images generally increases as the shear parameter increases. In low-shear systems, there is a strong correlation between image separations and the shear values, whereas it is reversed for the high shear cases, as shown in the lower panel in *Figure* 1. It is also interesting to note that the major difference between the mean scales of 2-image separations remain near the value of 2.0 in a unit of Einstein radius for all low shear systems, whereas its mean value is less than 1.0 in high-shear systems. We therefore notice that the high-shear lensing produces smaller observable scales, even though its external tidal perturbation is larger. For the low-shear case, there is a consistency where the maximum separations of 2-image lensing systems and of 4-image lensing systems are located within almost the same ranges.

It is well known that the maximum number of images for the *SIS*-lens system under a low-shear is 4 (*see Figure 5* in Section 3) when a source is located inside a tangential diamond caustic. In contrast, when the shear is larger than the critical value of 1.0, we found that the maximum number of multiple images is 3 (*see Figure 6* in Section 3). One of the faint images appears nearby the central region of a lensing galaxy, where the optical depth of the inter-galactic medium should be higher. As its magnification is usually quite low, it is just like the well-known central-image property of a non-singular lensing galaxy. However, if an extended source is assumed as usual in real galaxy-lensing, the third image may appear to be a comet like shape by connecting the other images (*see Figure 6* in Section 3). Notice that such exotic features of multiple imaging in the high-shear galaxy-lensing have not been reported in literature so far.

In *Figure* 2, the probability distribution for the individual magnification of the *SIS + shear* model is shown. Their general differences between the low- and the high-shear cases are shown in Figure 2. We note that as the shear value increases the corresponding magnification distribution becomes wider at both high- and low-magnification regions for the low-shear cases, but for the high-shear cases only the low-magnification region becomes dominant with increasing the shear.

In *Figure* 3 we present the distribution of magnification ratios (*hereafter*, *M.R*) between the fainter ($m_2$)- and the brighter-image ($m_1$) pairs of 2-image lensing systems. We defined the term as *M.R,* so that the maximum is a unity (i.e., *M.R* = $m_2/m_1$ ; $m_1 > m_2$). For the low-shear cases (< 1.0), the slope of distribution generally gets steeper when the shear value increases, which means that the highly contrasted magnification of the 2-images are dominant. On the other hand, for the cases of high-shear (> 1.0), there are in general flat distributions of magnification, but the narrow little peaks around the unity magnification ratio (i.e., *M.R ~ 1.0*) show that equally magnified 2-images are dominant. Therefore, we note that there are such distinctive differences of *M.R* of 2-image lensing systems between the low- and the high- shear cases.

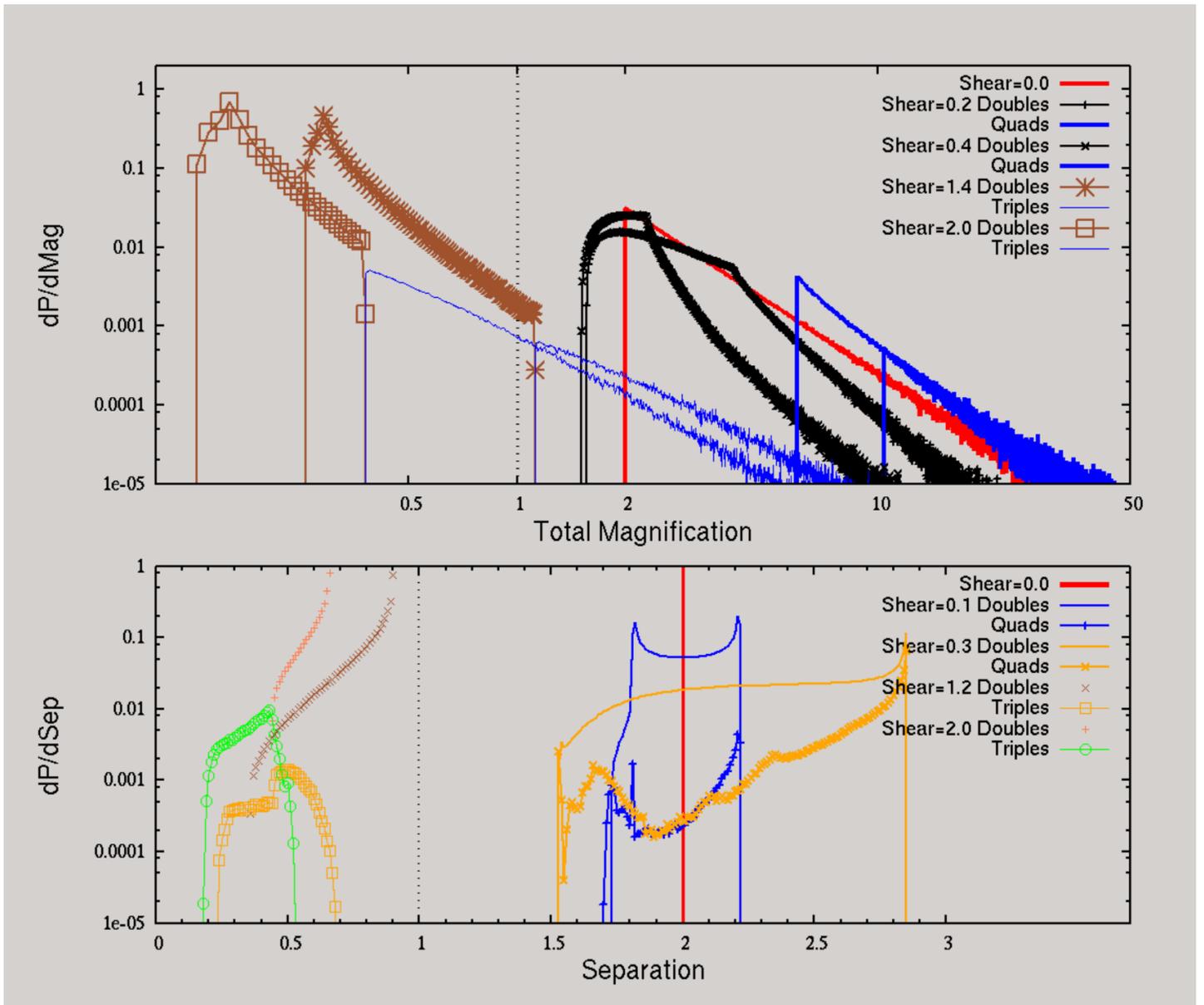

**Figure 1.** The imaging properties of the 2- and the 4-image lensing in the *SIS + shear* model. Top panel: total magnification probability distributions (dP/dM$_{tot}$) as a function of magnification (M) with the external shear (i.e., 0, 0.2, 0.4, 1.4, & 2.0). Bottom panel: the probability distributions of maximum image separations (dP/dSep) in normalized units with the external shear (i.e., 0, 0.1, 0.3, 1.2, & 2.0). The color represents corresponding shear values as described in the captions inside. The legends inside boxes represent the shear and the image multiplicity. Note the distinctive features of two-different shear groups in each panel; the left side represents high-shear cases and the right side represents low shear cases. The dotted vertical black line shows a unit magnification (upper-panel) or a unit separation (lower-panel).

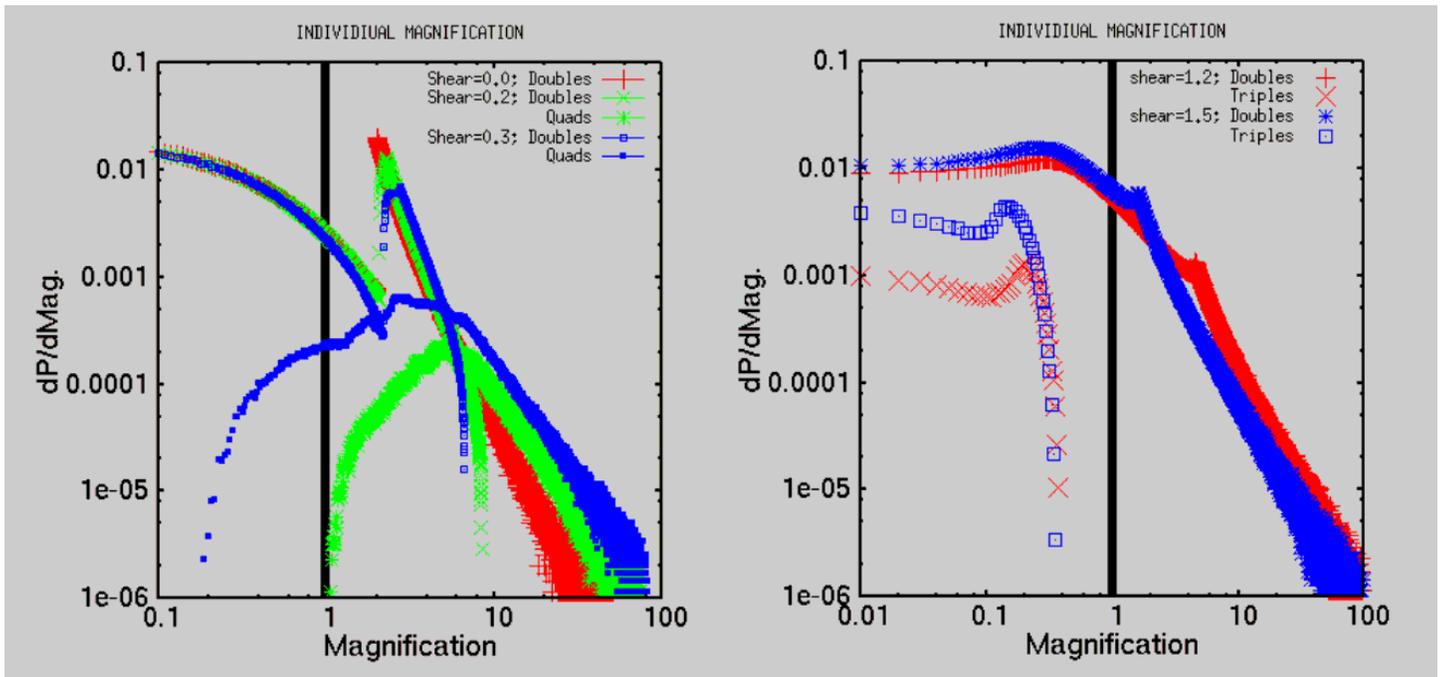

**Figure 2.** Individual magnification probability distributions (dP/dM$_{ind}$) of low (left)- and high (right)-shear cases. Left-panel: Note that the zero-shear case has only 2-image lensing, whereas the other low-shear cases (i.e. 0, 0.2, & 0.3) have both 2-image lensing and 4-image lensing. Right-panel: The two high-shear cases (i.e., 1.2, & 1.5) are shown with the 2-image lensing and 3-image lensing (i.e., the maximum image number in the high shear cases is 3). The legends inside the boxes represent the shear parameters and image multiplicity. The vertical black lines show a unity of magnification.

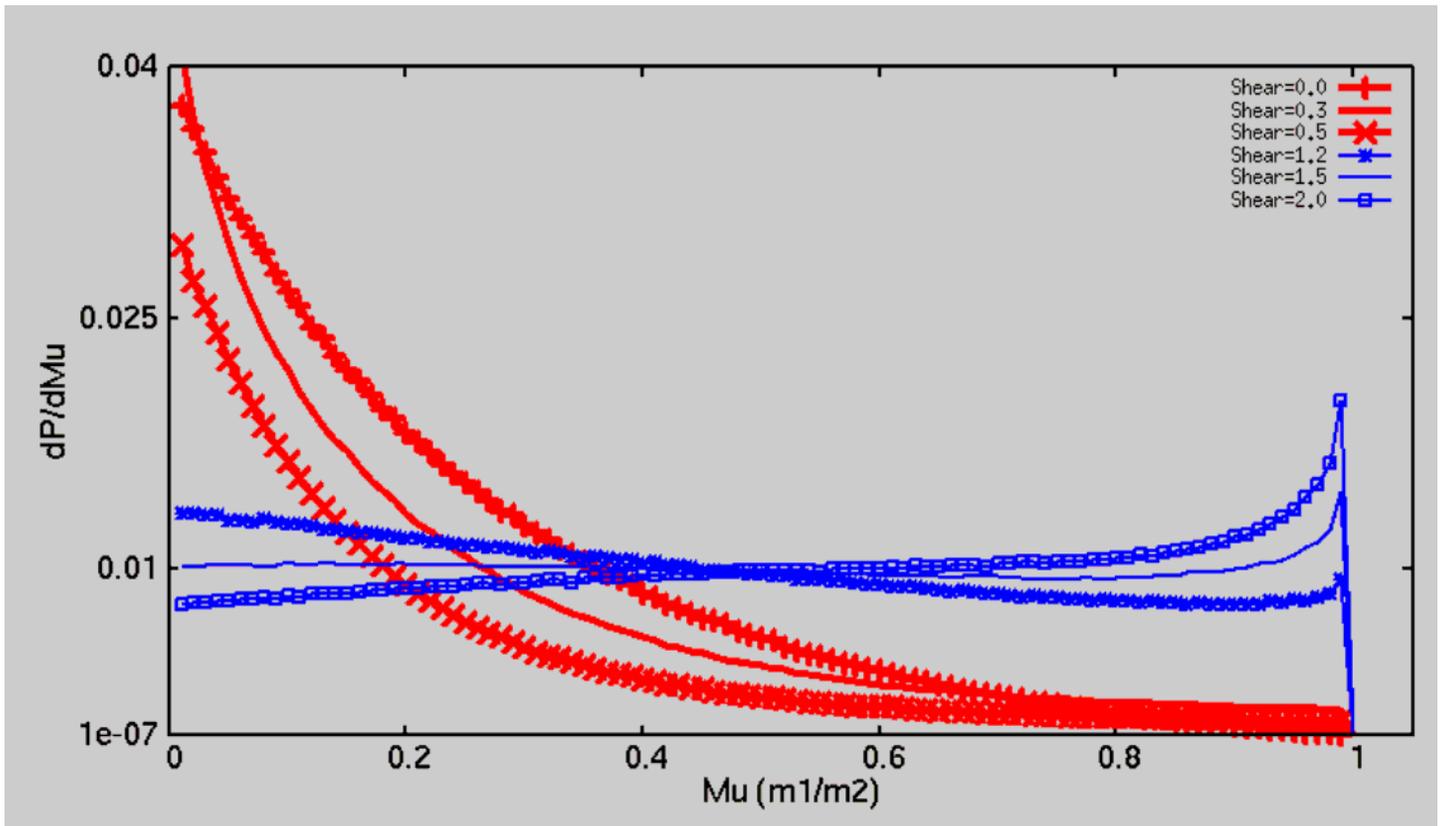

**Figure 3.** Magnification ratio ($M_u$ = m2/ m1) probability distributions of 2-image lensing systems as a function of shear. The shear values are 0, 0.3, & 0.5 (low-shear cases; <1.0) vs. 1.2, 1.5, & 2.0 (high-shear cases; >1.0). The legends inside the box represent the adopted shear parameter. The distributions of low-shear cases show an anti-correlation between the shear and the magnification ratios of 2-image lensing, while those of high-shear cases have flatter distributions and have a narrow little concentration near the unity of the magnification ratio.

## 3. Lensed Images with Caustics

### 3.1 Critical curves and Caustics

Various configurations of the caustic and multiple images of *SIS + shear* lens models are illustrated as a function of shear for a range between 0.0 and 2.0. The pure *SIS* model has no tangential caustic and only the *cut* instead (i.e. pseudo circular caustic; Schneider et al.1993, and reference therein), which is identical to the Einstein ring that is determined by the strength of a lensing potential. Note that the Einstein ring size remains unchanged in *SIS* models regardless of the adopted shear, but for the other galaxy lens models (i.e., *SIE*, tri-axial lens models) its size can be slightly changed depending on its ellipticity or tri-axial parameters (Schramm 1990; Kormann et al. 1993; Kassiola & Kovner 1993; Chae 2003; Keeton 2003; Rozo, Chen, & Zentner 2013), which is the so called *dynamical normalization* (e.g. Chae 2003) in lensing calculations.

If the external shear is added on the *SIS-lens* model, a central diamond caustic is generated where 4-image lensing occurs. In *Figure* 8, we illustrate the variations of critical curves and corresponding caustic curves of the *SIS + shear* model for both the low- and high-shear conditions. While the shear is less than 0.33, the diamond caustic is always smaller than its outer pseudo circular caustic curve (i.e., hereafter, *cut*), but after passing over the specific value of shear (~0.33), the diamond caustics transgress the size of the *cut*.

Note that the caustic in the source plane is dramatically increased as the shear becomes stronger. While radial caustic is a single curve for all the low-shear cases (< 1.0; *see Figure 5*), for the high-shear cases (> 1.0; *see Figure 6*), it becomes a complex curve accompanying the self-intersections and triangular caustics, as similarly predicted for the *Chang-Refsdal* lens model under a high shear (Chang & Refsdal 1979; Chang 1981).

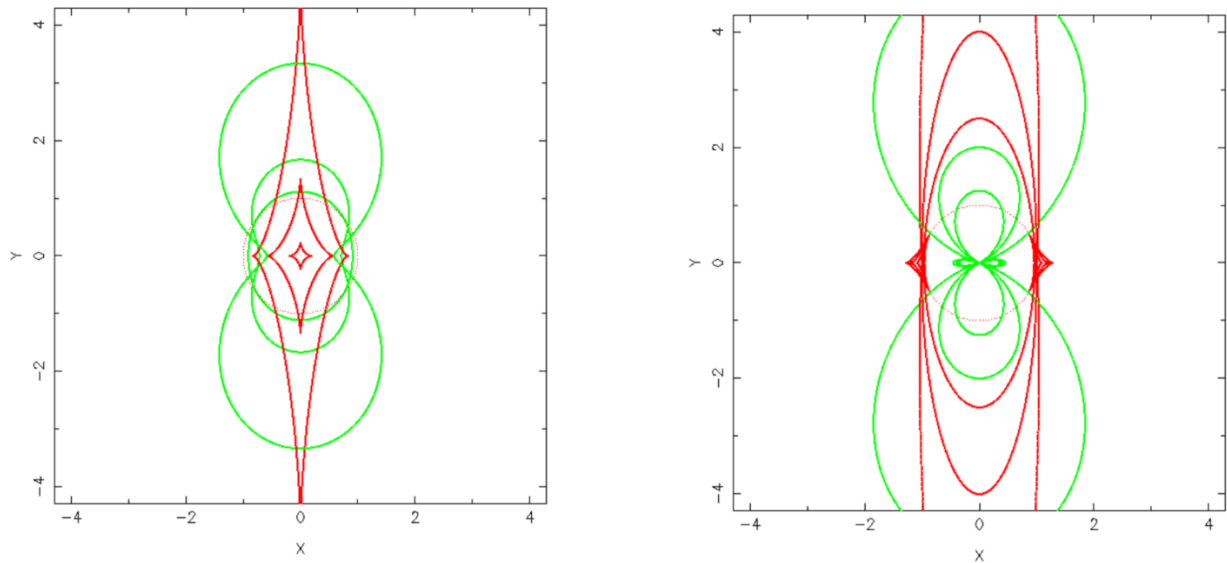

**Figure 4.** Caustics of the *SIS + shear* lens for the low- and high-shear cases. Left panel: the low-shear caustics with the shear of 0.1, 0.4, and 0.7, respectively. Right panel: the high-shear caustics with a shear of 1.2, 1.5, and 1.8, respectively. The green lines are critical curves in the lens plane, and the red lines are caustics in the source plane. The circular red dotted line represents a pseudo-caustic, the so called, *cut*. Note the variation of caustics heavily depends on its shear; it becomes extremely elongated when the magnitude of shear approaches a unity.

## 3.2 Multiple Images

We illustrate several examples of multiple lensed images for two low-shear cases (i.e., 0.0 and 0.2) shown in *Figure 5* and for a high-shear case (i.e., 2.0) shown in *Figure 6*, which are generated by our own interactive graphical-lensing software, *the Art of Photons (Lee 2003)*. In general, the *SIS + shear* model produces the maximum of 4 images under a low shear condition (i.e., <1.0). However, in a high shear condition (i.e., >1.0), we found that the maximum number of images is 3. As for the property of multiple lensed images, high shear models produce significantly different image configurations in terms of the maximum number of images and their magnification. The multiple images under the high-shear conditions usually have lower magnification than the low-shear conditions, except for the two small regions of triangular shaped caustics in which 3-image lensing can occur (*see* the last two rows in *Figure* 6).

As shown in *Figure* 6, we only used a single high-shear of 2.0 as the shape of the high-shear caustic did not change so much by the adopted shear value except for their scales. Note that the multiple imaging properties of the high-shear cases shown in *Figure* 6 are quite different in their characteristics from those of the low-shear cases shown in *Figure* 5. Some exotic shapes (e.g. like water drops) of multiple images can be generated for the cases of an extended source.

Furthermore, for the high-shear cases, we interestingly found that the parities of 3-image lensing systems always produce the same negative parities, but 2-image lensing systems can produce an image pair with either same parities or different parities. Since under the low-shear cases the parities of 2-image lensing systems always have different parities, this is the most distinctive imaging feature of the 2-image lensing systems of a high-shear condition.

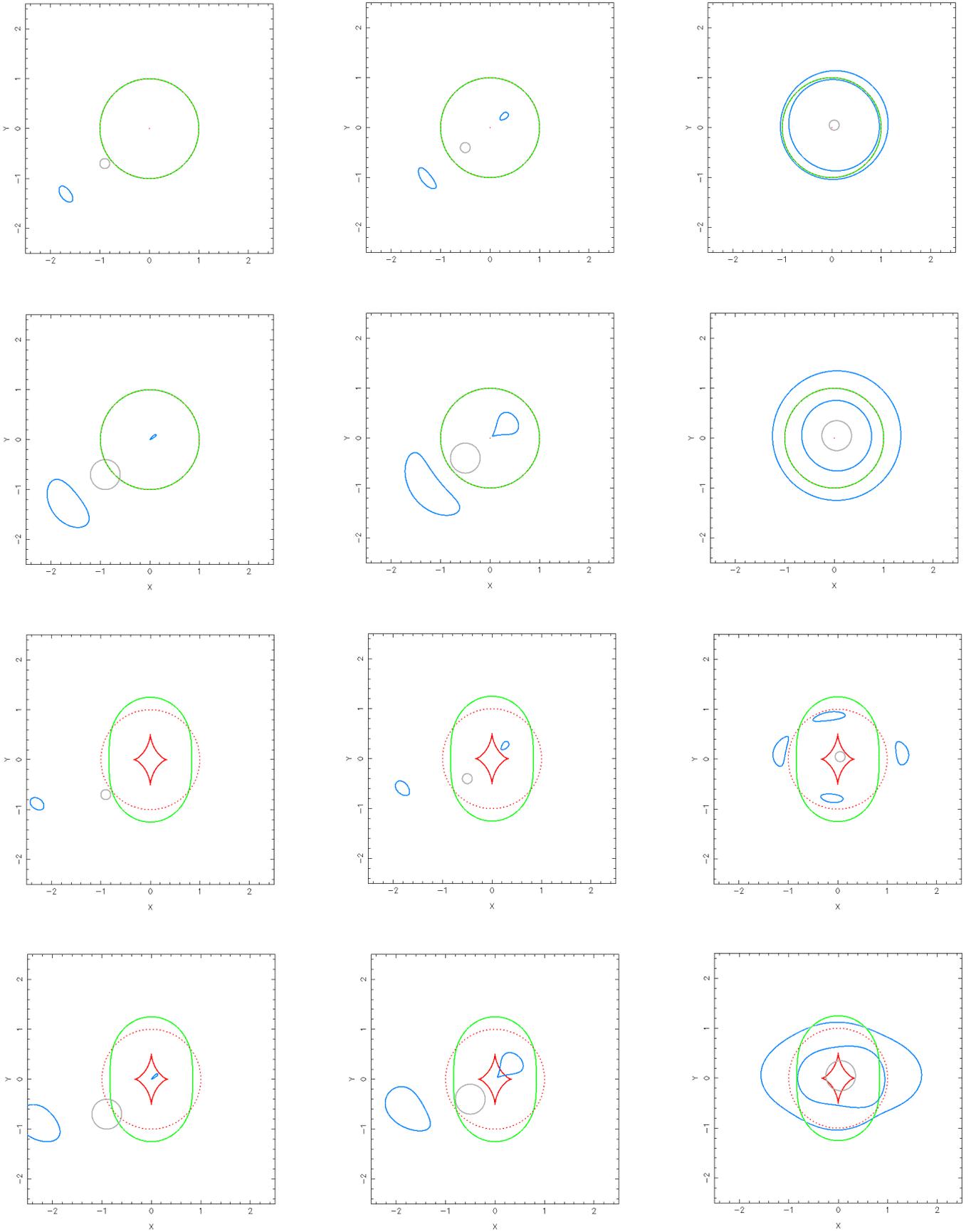

**Figure 5.** Multiple images of the *SIS + shear* model with low shears (i.e., shear < 1.0). The first two rows show zero shear cases and the second two rows are for the cases of 0.2. Two different source radii are used: a small source of 0.05 and a large source of 0.3, and their positions are (-0.9:-0.7); (-0.5:-0.4); (+0.05:+0.05) in a unit of Einstein radius for each column. Note that the boundaries of multiple images are accurately generated by equation 2 and the maximum number of images is 4. The circular source is in gray and the corresponding multiple images are in blue. The green lines are critical curves, dotted circle is the *cut*, and the red lines are caustics.

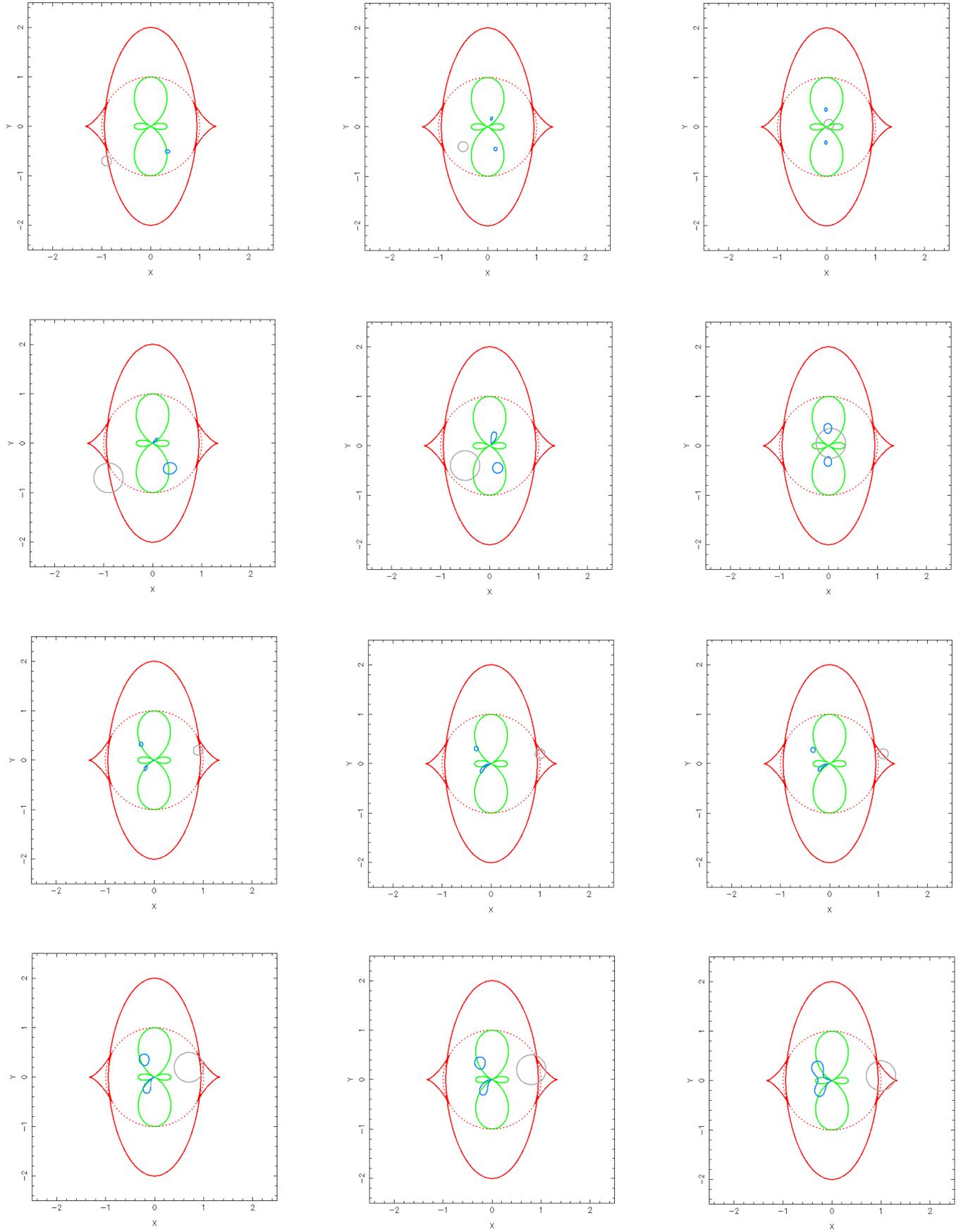

**Figure 6.** Multiple images of the *SIS + shear* model with a high shear (i.e., shear > 1.0). The shear is fixed to 2.0 for all cases. Note that an oval shape caustic and two small concaved triangular caustics appear and the maximum number of images is 3. The two different source sizes and the colors are the same as those in Figure 5. The last two rows present specific imaging properties of small concaved-triangular caustics in which 3-image lensing occurs.

# 4. Lensing Cross-sections

## 4.1 Magnification Bias

In gravitational lensing, all of the important multiple-imaging parameters (i.e., caustics and source) are actually located in the projected source-plane, while a lensing object is located alone in the lens-plane. Since the source's position, with respect to caustic, determines its image multiplicity and magnifications, the lensing cross-section for each image multiplicity is directly related with the caustic's property. As the lensing cross-sections are proportional to the size of the caustic in the source-plane where multiple images occur, we should integrate appropriate caustics in the source plane depending on image multiplicity. Based on the inverse mapping formula shown in equation 2, we carry out intensive Monte-Carlo simulations in order to find corresponding caustic areas depending on image multiplicity.

Finch et al. (2002) made a semi-analytic analysis on the cross-section of *SIS + shear* model, in which they suggested approximation formulas for the variation of the cross-section and magnification relations as a function of shear value. However, these formulas are valid only with a relatively low shear condition (i.e., < 0.3) as they stated. We also note that they made use of a different method in calculating magnification effects by using the general relation of extended areas between the source- and the lens-planes. In their study, they made theoretical model calculations for the lensing cross-sections, however, without considering the reducing effect of the 2-image lensing areas due to the observational survey selection bias.

However, we notice that the cross-sections of 2-image lensing are also affected by the given survey selection limit limits as shown in *Figure 7*. This implies the detection ability of the contrasted flux levels between a primary (i.e., brighter) image and a secondary (i.e., fainter) image. Also, the magnification bias calculation should also consider the differential cross-section effect as their sizes are changed by a given value of the survey selection limit.

One can notice the fact that when 2-image lensing occurs with a source located at the outer-edge region inside the Einstein ring, it only produces highly-contrasted 2-images as shown in *Figure* 3 (refer to the cases for the low-shear cases). Thus, such a specific outer area producing highly contrast 2-images should be removed in calculating both the cross-section of 2-image lensing and its corresponding magnification bias effect as Mitchell et al. (2005) properly made use of this effect in their lensing calculations. However, note that Mitchell et al. (2005) used a pure *SIS* model case without considering the external shear parameter, like the most of previous N-body galaxy-lensing simulations. Usually, the pure *SIS* without shear model has been widely used in N-body galaxy lens simulations (e.g. Katz & Paczynski 1985; Claeskens et al. 2001) by far.

To calculate the areas of cross-section and each individual magnification of a point source, we carried out intensive Monte-Carlo simulations by using the inverse mapping formula in equation 2. In general the magnification bias and the biased cross-section can be described as the following relations as shown in equation 4 (e.g. Turner et al.1984; Huterer et al. 2004; Mitchell et al. 2005), which combines the source luminosity function and magnification probability distribution with a magnification bias. The magnification bias and the *biased* cross-section (e.g. Mitchell et al. 2005) can be conventionally defined as follows,

$$B_{img} = \frac{\int_{Smin}^{\infty} dS \int_{\mu min}^{\infty} \mu^{-1} P_{img}(\mu) N\left(\frac{S}{\mu}, z\right) d\mu}{\int_{Smin}^{\infty} N(S, z) dS}, \qquad (5)$$

$$\sigma_{biased} \equiv A_{img} \cdot B_{img}.$$

where, $B_{img}$ is the magnification bias factor of each image multiplicity (*img*), $P_{img}(\mu)$ is the distribution of total magnification, $\mu_{img}$ is the total magnification for image multiplicity, $\sigma_{biased}$ is the so called, *biased* cross-section depending on the image multiplicity, which is the product of the cross-section area ($A_{img}$) and magnification bias ($B_{img}$), N(S,z) is the source's luminosity function with the source luminosity (*S*), and the redshift (*z*).

For the case of a simple power-law luminosity function of sources, its magnification bias can be reduced to $<\mu_{img}^{\beta-1}>$ (Rusin & Tegmark 2001; Finch et al. 2002), where, $\beta$ is the power index of the source's luminosity function. The magnification bias represents the fact that faint sources can be observed over an observational threshold due to the lensing magnification, which is a combination of luminosity function and magnification probability distributions of image multiplicities. Any observational surveys, whether it is an optical-survey or radio-survey, are flux-limited in sampling, which set a limit on the observational threshold between a brighter- and a fainter-image of 2-image lensing systems. Then, if the survey selection limit is not enough to detect a severely contrasted 2-image lensing system, it cannot be identified as a 2-image lens system.

We note that there was so far no extensive analysis on the lensing cross-section for a wide range of shears beyond the extreme shear realm over 1.0. In this work as our aim goes further by including a wide range of external shears, we systematically investigate their effects on the lensing cross-section.

Here, we define the following two conceptual terms in our work as shown in equation 5: *the differential lensing* cross-section and *the effective lensing* cross-section for the 2-image lensing, 3-image lensing, and the 4-image lensing. *The differential lensing* cross-section only accounts for the effect of survey selection limit with the external shear, and *the effective lensing* cross-section accounts for the final convolved effect by *the differential lensing* cross-section and its magnification bias effect depending on each *differential* cross-section.

Notice that the magnification bias effect also depends on its *differential lensing* cross-section as its size depends on a given survey selection limit. Depending on the adopted survey selection limit and the shear, its *differential lensing* cross-section and its corresponding average magnification were calculated by Monte-Carlo methods. We compute the lensing cross-section and the magnification bias separately. Since it is essential to correctly take into account *the differential* cross-section areas and its corresponding magnification bias value as mentioned above; the magnification bias directly depends on the *differential* cross-section pre-determined by a survey selection limit in observations as shown in *Figures* 7 and 8.

Then, we define the final term, *the effective lensing* cross-section, based on *the differential lensing* cross-section and the normalized magnification bias depending on a survey selection limit. These numerical terms in our calculation method are defined as follows,

$$\sigma_{multi} = \left(\sigma_2^{Dif} + \sigma_3 + \sigma_4\right),$$

$$\text{where, } \sigma_2^{Dif}(f_{limit}) = \left(\pi R_{Ein}^2 - \sigma_4 - \sigma_{f_{limit}}\right), \qquad \sigma_3^{Dif} = \sigma_3, \qquad \sigma_4^{Dif} = \sigma_4,$$

$$\sigma_{Doubles}^{Eff} = \sigma_2^{Dif} \cdot B_n, \qquad \sigma_{Triples}^{Eff} = \sigma_3^{Dif} \cdot B_n, \qquad \sigma_{Quads}^{Eff} = \sigma_4^{Dif} \cdot B_n, \qquad (6)$$

$$\text{where, } B_n(f_{limit}, \beta) = \left(\frac{B_{img}}{\pi R_{Ein}^2}\right),$$

$$\left(\frac{N_4}{N_2}\right) = \frac{\sigma_{Quads}^{Eff}}{\sigma_{Doubles}^{Eff}}, \qquad \left(\frac{N_3}{N_2}\right) = \frac{\sigma_{Triples}^{Eff}}{\sigma_{Doubles}^{Eff}}.$$

where, $\sigma_{multi}$ is the total cross-section of multiple-lensing. The *differential lensing* cross-sections that correspond to each image's multiplicity are denoted by $\sigma_2^{Dif}$, $\sigma_3^{Dif}$, *and* $\sigma_4^{Dif}$, respectively. $\sigma_{f_{limit}}$ is the outer-edge cross-section corresponding to the area producing highly contrast 2-images below the survey selection limit. *The effective lensing* cross-section, which is convoluted by the proper magnification bias for doubles, triples, and quads, are denoted by $\sigma_{Doubles}^{Eff}$, $\sigma_{Triples}^{Eff}$, and $\sigma_{Quads}^{Eff}$, respectively. $R_{Ein}$ is the normalized Einstein radius, $B_n$ is the magnification bias factor normalized by the Einstein ring area as a function of the survey selection limit (e.g., *$f_{limit}$*= 0.1, for the case of *CLASS*; Chae 2003) and β is the power index of the source's luminosity function. $N_4$, $N_3$, and, $N_2$ represent the final ratios of the *effective cross-sections* in each image's multiplicity.

Throughout this work, we adopt a fixed power-index (β) of 2.0 (e.g. Rusin & Tegmark 2001) for a simple power-law luminosity function (i.e., $N(S) \propto S^{-\beta}$) of the flat spectrum radio sources. This value represents the generally obtained values from two major radio-lens surveys, *JVAS/CLASS* (e.g. Patnaick 1998; King & Browne 1996).

**4.3 Lensing Cross-sections**

Adopting the above relations into our Monte-Carlo scheme, we carried out intensive calculations over a wide range of shear parameters while considering several predefined values of survey selection limits (i.e., $f_{limit}$ = m$_2$/m$_1$; *hereafter, f.l*) between a fainter image magnification (m$_2$) and a brighter image magnification (m$_1$). This survey selection limit actually depends on the dynamic range of each survey in different wavelengths; we adopt it to be 0.1 (e.g. Chae 2003) for the *CLASS* radio survey.

In *Figure* 7 we illustrate *the differential lensing* cross-sections in the source-plane with various survey selection limits and the adopted shear. Several different colors in the 2-image lensing region clearly illustrate the dependence of arbitrary limits on the survey selection limit (f.l) with their proper cross-sections; the 3-image lensing regions are the white areas in the naked cusps which transgress over the circular pseudo caustic; 4-image lensing region is the internal diamond tangential caustic.

Finch et al. (2002) suggested some useful approximation relations for the cross-section of 4-image lensing (i.e., $\gamma^2/(1-\gamma^2)$) and its mean magnification (i.e., $\gamma/(1-\gamma^2)$) of the *SIS + shear* model. Huterer et al. (2004) also investigated the effects of the ellipticity and of the shear on the cross-section variations. However, it is worth to note some features of the previous studies for the cross-section of lensing galaxies. Finch et al. (2002) made a pure theoretical analysis on the cross-sections, but did not consider the additional effect of survey selection limits. Huterer et al. (2004) also made a similar theoretical study, but did not consider such an internally correlated variation in accordance with the survey selection limits when calculating the cross-section of image multiplicity and its corresponding magnification bias. Their studies are correct in a pure theoretical realm. However, in reality, we note that one must consider these observational side-effects being aroused from the given limited observational selection bias, which is determined by the observational dynamic ranges in surveys. However, Mitchell et al. (2005) properly considered the reduction of the cross-section of 2-image lensing adopting the survey selection limit of 0.1 in *CLASS*.

As this kind of survey selection limits are different for wavelengths at which observations are carried out, there is a big difference between optical observations and radio observations. For example, the HST optical observations can easily discriminate double images having a flux difference of 1/100 between a bight image and a faint image. However, in radio observations, such flux discriminating ability for double images is much lower than optical observations, as the *CLASS/JVAS* surveys have a limit around 1/10 ~ 1/20 (King et al. 1999; Chae 2003).

Thus, we also considered this reducing effect on the 2-image lensing thoroughly in the wide range of the shear, when calculating differential cross-sections and their corresponding magnification bias. We must notice that not only the shear, but also the survey selection limit determines the size of the cross-section, and affects its corresponding magnification bias, especially for the double lensing cases. Thus, the shear and the survey selection limit should be considered simultaneously in calculating the cross-section to account for the observed frequency of multiple images.

We found that this is the major reason for having different estimates on the probable shear value for *the high Quads-to-Doubles ratio*. Regardless of the detail of lensing galaxy models (e.g. ellipticity, shear, or tri-axiallity), this survey selection limit should affect any other galaxy-lens models since it is actually an observational by-product depending on the current technology for different wave-length surveys. Such a reduction effect due to the flux magnification ratio was also considered by King et al. (1999), but they just assumed a simple *SIS* model without the shear and estimated about 25% reduction of lensing events in the *JVAS*. However, if the external shear (or internal ellipticity) is included, its relative reduction becomes more complex as shown in *Figure 7*.

Also, while the external shear increases, the corresponding *differential lensing* cross-section preferentially suppresses the cross-section of the 2-image lensing systems, but the cross-section of the 4-image lensing systems increase as shown in *Figure* 7. Such a strong dependence of 2-image systems on the survey selection limit is illustrated in the 4 exemplary shear cases (i.e., *0, 0.2, 0.5, and 2.0*). Note that such a differential effect on the 2-image lensing cross-sections is determined by a given survey selection limit and the shear parameter. The total lensing cross-section without the shear effect is equal to the area of the Einstein ring.

In *Figure* 8, the variation of cross-sections depending on several survey selection limits (i.e., 0.0, 0.01, 0.05, & 0.1) are presented as a function of the shear ranging from 0 to 2.0. The *Quads-to-Doubles ratio* becomes a unity when the shear is 0.55 for the case of total cross-section (i.e., f.l = 0.0) of 2-image lensing. However, it is noticeable that the total 2-image lensing area decreases dramatically down to 67% for the case with a survey selection limit of 0.1. There is a strong anti-correlation between the cross-sections of the doubles and of the quads, since the size of a tangential (diamond) caustic internally trespasses the size of the radial caustic when the shear increases. The survey selection limit always preferentially reduces the cross-section of 2-image lensing. Therefore, there is a systematic loss in 2-image lensing cross-sections which strongly depend on both the shear and the survey selection limit, while the 4-image lensing cross-section, however, depends only on the adopted shear as there is no loss of cross-section due to the survey selection limit.

In *Figure* 9 we present the estimated *Quads-to-Doubles ratio* and the *Triples-to-Doubles ratio* as a function of the shear based only on the *differential lensing* cross-sections shown in equation 5. Depending on the survey selection limit, the corresponding shear varies to the same *Quads-to-Doubles ratio* and the *Triples-to-Doubles ratio*. We note that the larger the survey selection limits, the less the shears can achieve the same *Quads-to-Doubles ratio*.

Depending on the several different limits of the survey selection limit, 30% of the *Quads-to-Doubles ratio* occurs at the shear value of 0.24, 0.29, 0.35, and 0.37, respectively. Also, 1% of the *Triples-to-Doubles ratio* occurs at the shear values of 0.40, 0.44, 0.45, and 0.48, respectively, for the low-shear region (i.e., < 1.0). But, for the high-shear region (i.e., > 1.0), it occurs only at the shear of 1.2.

In *Figure* 10 we present the estimated final term in equation 5, which is *the effective lensing* cross-section as a function of shear that convolves the above *differential lensing* cross-section with its magnification bias effect of the source's luminosity function. For the case of the high-shear realm, *the effective lensing* cross-section is anti-correlated with the magnitude of shear as their magnification probability distributions are also anti-correlated with the shear as shown in *Figure* 1. We note that if the survey selection limit is assumed to be 0.1, 30%, 50%, and 70% of the *Quads-to-Doubles ratios* are expected to occur at the shear value of 0.13, 0.16, and 0.183, respectively. Therefore, the *high Quads-to-Doubles ratio* of 50% ~ 70%, observed by *JVAS/CLASS* radio-lens surveys, can be explained by the moderate shear range of 0.16 ~ 0.18. We notice that this new shear estimation is about half of the previous estimate of 0.3 (Finch et al. 2002; Huterer et al. 2004).

One may compare the results of *the effective lensing* cross-section in *Figure* 10 with the results of *Figure*s 8 and 9, which only considered *the differential lensing* cross-section without any additional magnification bias. Notice that the estimated ratios of *the Quads-to-Doubles* and *the Triples-to-Doubles* increase more steeply when the shear and the survey selection limit increase, especially for the cases of *the effective lensing* cross-section shown in *Figure* 10.

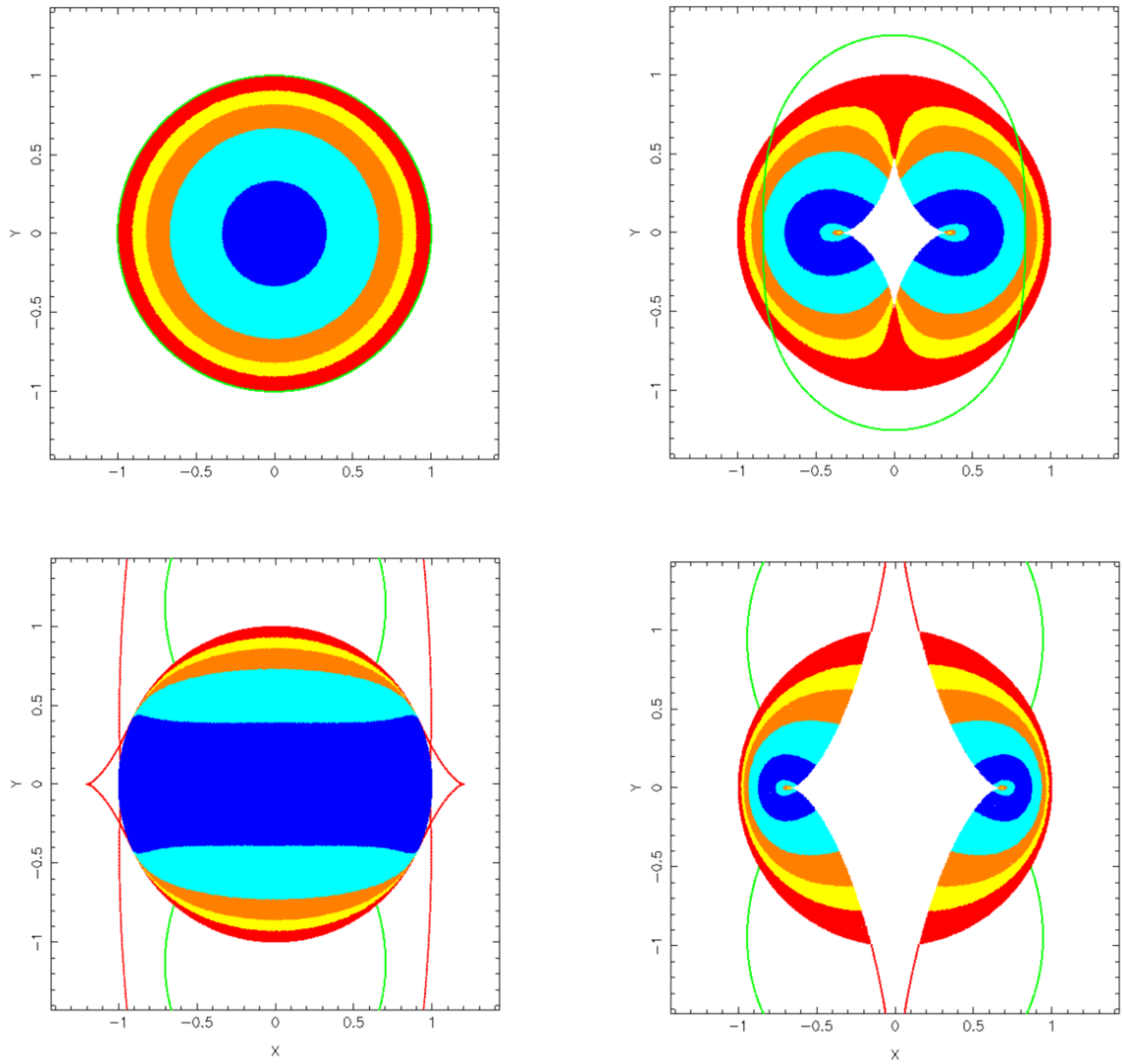

**Figure 7.** Various differential lensing cross-sections of the 2-image lensing systems in the *SIS + shear* model. From the top left, clock-wise: zero-shear case; the case for 0.2 (cf. the central diamond caustic is for 4-image lensing); the case for 0.5 (cf. the naked cusps transgress the radial caustic, where triples occur); and the case for a high shear of 1.5. The various colors in each panel show differential cross-sections of 2-image lensing systems corresponding to several limits of the survey selection limit (i.e., blue: f.l > 0.5; sky: f.l > 0.2; dark yellow: f.l > 0.1; yellow: f.l > 0.05; red: f.l > 0.01). The critical curve is the green line and the caustic is the red line. Note that the cross-section of 2-image lensing strongly depends on both values of the survey selection limit and the shear.

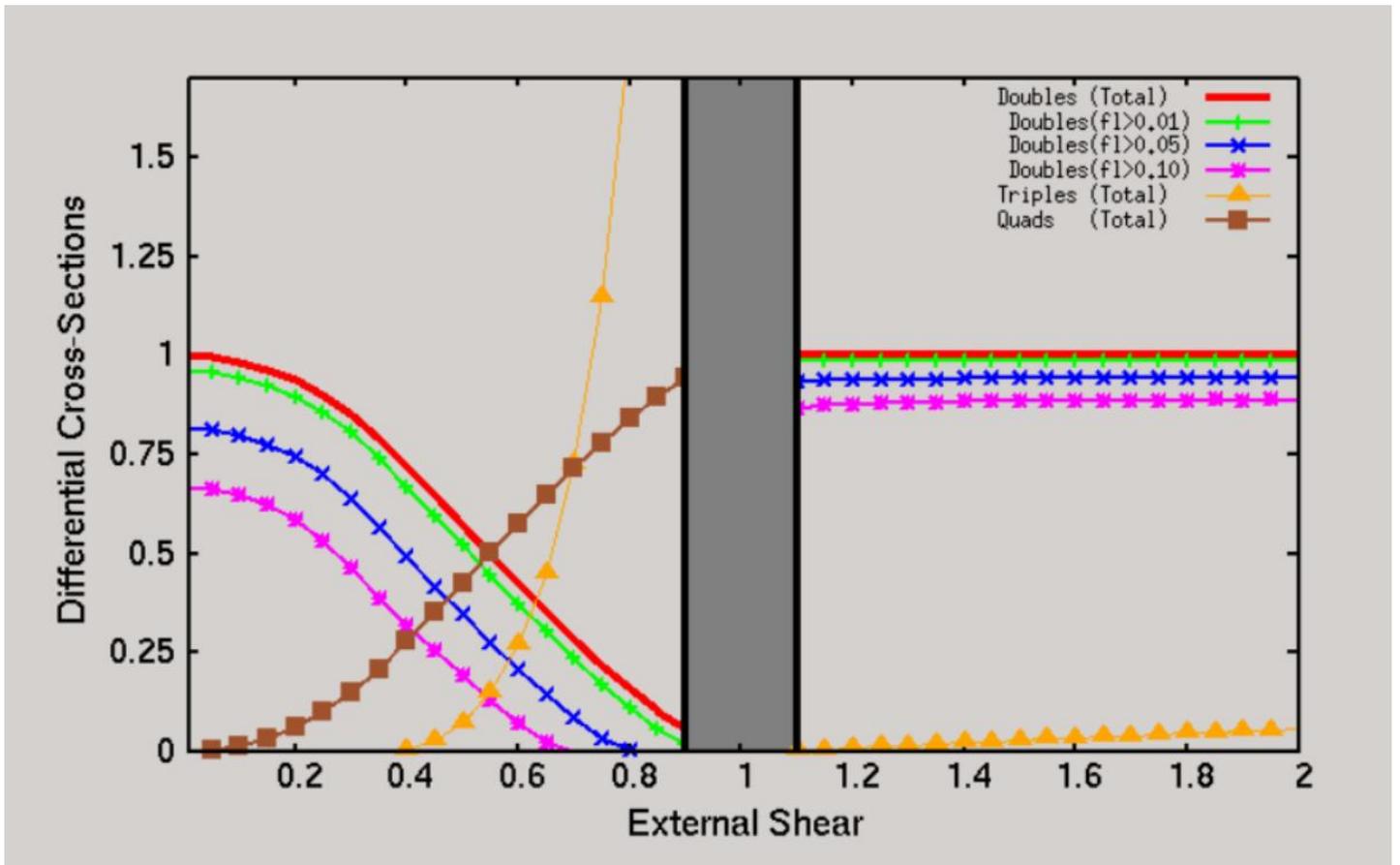

**Figure 8.** The differential lensing cross-sections of the *SIS + shear* model as a function of shear (0.0 ~ 2.0). The cross-sections of 2-image lensing for several predefined limits of survey selection limit (i.e., f.l = 0.0, 0.01, 0.05, and 0.1), the cross-section of 3-image lensing (triangle), and that of 4-image lensing (square) are shown. The legends inside the box represent the image multiplicity and the survey selection limits (f.l). The blank area in the middle of the X-axis (between 0.9 ~ 1.1) is not considered here, because of its extremely elongated caustic. The values are normalized by the cross-section of the Einstein ring. Note that the survey selection limit is valid only for the 2-image lensing systems.

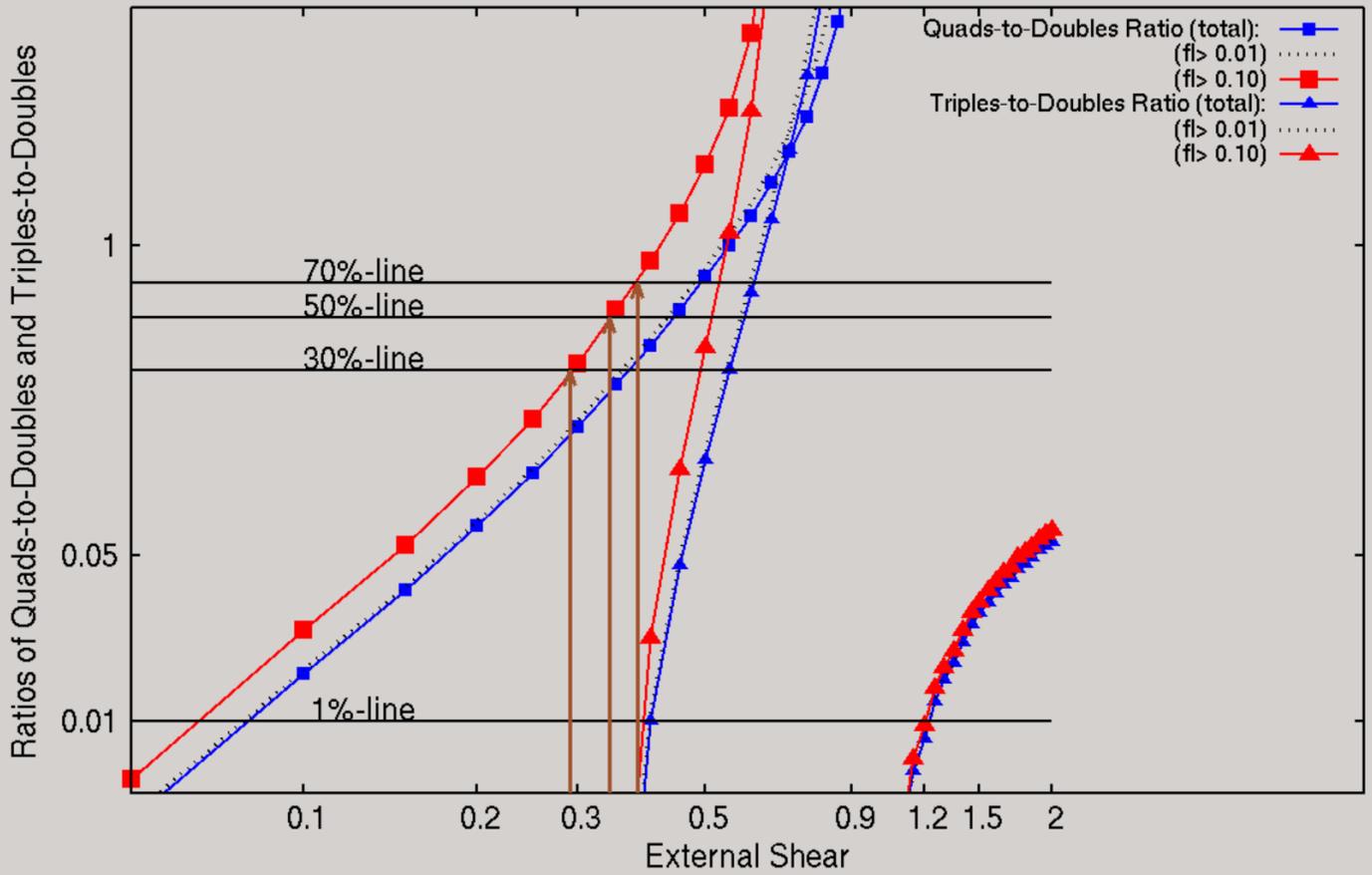

**Figure 9.** The ratios of Quads-to-Doubles and Triples-to-Doubles based on the differential lensing cross-section. The legends inside the box represent the image's multiplicity (symbols) and the different survey selection limits (i.e., f.l = 0.0, 0.01, and 0.1). The three vertical arrows indicate the 30%, 50%, and 70% of the Quads-to-Doubles ratios occurring at the shear of 0.29, 0.33, and 0.38 if the survey selection limit is set to 0.1. 1% of the Triples-to-Doubles ratio occurs around 0.4 and 1.3 of the shear axis.

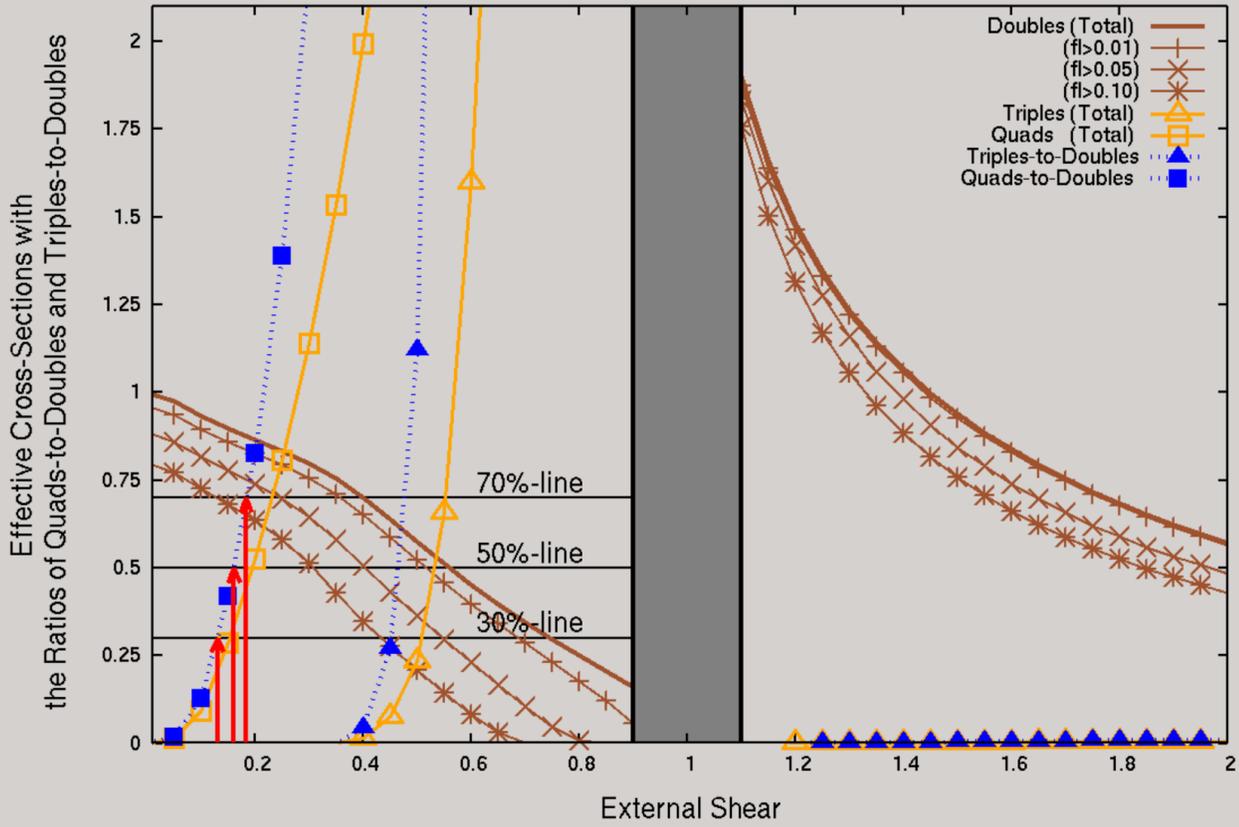

**Figure 10.** The effective lensing cross-sections of the *SIS + shear* model as a function of shear (0.0 ~ 2.0). This effective cross-section is the combination of the differential lensing cross-section and the magnification bias of the radio source's power law luminosity function with a fixed power-index (i.e., β=2.0). The legends inside the box represent the image's multiplicity and different survey selection limits (i.e., f.l = 0.0, 0.01, 0.05, and 0.1). The two symbolized dotted-lines represent the finally estimated ratios of the Quads-to-Doubles (square) and the Triples-to-Doubles (triangle) if adopting the survey selection limit (f.l) of 0.1, respectively. Three red arrows indicate 30%, 50%, and 70% of the Quads-to-Doubles ratio with the f.l of 0.1(i.e., corresponding to *CLASS*), at which these shear values are 0.13, 0.16, and 0.183, respectively.

# 5. Summary and Conclusions

Based on a representative *SIS + shear* model, we present various results of lensing properties for a wide range of external shear perturbations ranging from 0.0 ~ 2.0. Based on this galaxy-lens model, we calculated its imaging and caustic properties for both the low-shear (< 1.0), and the high-shear (> 1.0) cases using Monte-Carlo simulations. Using the direct inverse mapping formula of the *SIS + shear* lens system, as presented in equation 2, we systematically calculated and compared the different features of total and individual magnification distributions, image separations, and magnification ratios of 2-image and 4-image lensing systems as a function of the external shear (*Figure*s 1, 2, & 3). It turns out that the shear also produces a trend in weakening the magnification regardless of the image's multiplicities. The detailed analysis of the lensing cross-section was also presented as a function of the external shear and the survey selection limit in Section 3.

The cross-section variations including the survey selection limit and magnification bias effects for 2-image and 4-image lensing cases, are numerically investigated for the shear range of 0.0 ~ 2.0. As for the long standing anomaly in gravitational lensed quasar surveys, such as the *high Quads-to-Doubles ratio* (e.g. King & Browne 1996; Cohn & Kochanek 2004; Bolton et al. 2006), our analysis on *the effective lensing* cross-section is finally presented in *Figure* 10.

This provides an insight in that there is a significant preferential reduction for *the effective lensing* cross-section of 2-image lensing systems. We clearly show that there is a significant loss on *the effective lensing* cross-section (*see* Section 4) of the 2-image lensing compared to the cases of the 4-image lensing. This is because of the strong dependence of the cross-section variation on the 2-image lensing systems, which is in accordance with the survey selection limits of different wavelength observations (*Figure*s 8, 9, & 10).

We also derived the expected lensing ratios of *Triples-to-Doubles* and *Quads-to-Doubles* as a function of the shear using several fixed limits of the survey selection limit. Since *the differential* cross-section affects significantly the reduction of the 2-image cross-section compared to 4-image lensing (*see Figure*s 7 & 8 in Section 4), such a difference in their *effective lensing* cross-sections which convolves *the differential lensing* cross-section (*see* Section 4) with a survey selection limit and its corresponding magnification bias, provides the main reason for this long standing problem (*see Figure* 10 in Section 4).

With a relatively low external shear value of 0.16 ~ 0.18, we thus found that the observed *high Quads-to-Doubles ratio* can be largely explained if the lensing cross-section and corresponding magnification bias are correctly considered in accordance with their survey selection limit. What could be the major difference between the optical- and the radio lens surveys? Although the radio survey has merits of angular resolution and interstellar reddening, the major difference is their different limits of the survey selection bias between a brighter image and a fainter image due to different dynamic ranges. Thus, discriminating the highly contrast 2-image lensing is an important factor in studying the properties of lensing frequency, since such a flux selection limit is much better in optical observations than in radio observations. Then, we are able to conclude that radio surveys may have missed more highly contrast 2-image lensing systems than optical surveys did, in spite of fine angular resolving power.

This conclusion is also very consistent with the N-body galaxy simulations of Cohn & Kochanek (2004) in which they stated that the presence of nearby satellite galaxies, which act like an additional shear to a main lensing galaxy, systematically suppresses the probability of obtaining 2-image lensing systems. Based on our cross-section analysis, it is directly understandable that the presence of nearby satellite galaxies does increase additional external shear effects to the main lensing galaxy, but systematically reduces the 2-image lensing. Finally, the 4-image lensing probability seems to be increases (*see* Figure 7, 8) than the 2-image lensing cases.

Holder & Schechter (2003) concluded that there is a likelihood of 45% on the possibility that the average shear produced by *LOS*-galaxies to a main lensing galaxy can be larger than 0.1 based on N-body galaxy simulations. Wong et al. (2011) similarly derived the possible shear contribution of the *LOS*-galaxies around the observed quasar lens systems to a main lensing galaxy of which the average magnitude can be ~ 0.08. We also note that the typical external shear from approximately ~ 100 multiply-lensed quasar systems (*refer to CASTLEs*) are usually affected by the order of 0.1 magnitude of the tidal shear field. Then, our estimate of the effective shear is consistent with previously anticipated values as our estimate is about a half of previous estimates (Finch et al. 2002; Huterer et al. 2004)

Therefore, our estimation of the averaged effective shear of 0.16 ~ 0.18 for most of the lensing galaxies, which can conceptually explain the anomaly of the *high Quads-to-Doubles ratio,* is also consistent with those previous N-body lensing studies (Holder & Schechter 2003; Cohn & Kochanek 2004; Wong et al. 2011). Since in reality any single lensing galaxy's potential is always affected by both an intrinsic ellipticity and the external shear, our shear estimation should be regarded as an averaged-effective shear term acting on a lensing galaxy, which implicitly combines both perturbation terms to the circular lensing potential.

As discussed by Rusin and Tegmark (2001), who studied the problem of *the high Quads-to-Doubles ratio*, they concluded that the required axial-ratio of *SIE* lens is about 0.4 (i.e., in terms of the shear magnitude ~ 0.13; Keeton, Kochanek & Seljack 1997) to explain the observed frequency. However, note that this only accounts for the high frequency of the 4-image lensing, not the observed low frequency of 2-image lensing. It seems that the authors may have missed the fact that there is a severe reduction of *the differential* cross-section of 2-image lensing systems compared with the case of 4-image lensing systems.

As any lensing galaxy has an intrinsic ellipticity in reality and there should be additional tidal perturbations produced by neighboring satellite galaxies and/or *LOS*-galaxies as concluded by Wong et al. (2011), our estimation for the most probable shear estimate of 0.16 ~ 0.18 seems to be very consistent with most of the previous estimates on the additional environmental shear. Note that our newly estimated most probable shear of 0.16 ~ 0.18 is about half of the previous estimation of 0.32 (e.g. Finch et al. 2002; Huterer et al. 2004), which used somewhat different approaches, as discussed in Section 4. Thus, it implies that much less environmental matter and dark matter around a lensing galaxy can explain the observed *high Quads-to-Doubles ratios,* as there is a certain degree of contributions from their internal ellipticity and tri-axiality. For example, the investigation of 11 nearby galaxy clusters by Jorgensen et al. (1995) found the mean ellipticity of 0.31.

Therefore, if adopting the representative mean ellipticity of 0.31 with *the shear and ellipticity relation* (i.e., $\gamma \sim \epsilon/3$; Keeton, Kochanek, & Seljack 1997), we can assume that our estimated total effective shear of 0.16~0.18 can be regarded as the combination of the internal ellipticity the real external shear in reality. Then, to interpret our shear estimate of 0.16 ~ 0.18, we assume that the mean internal shear is to be 0.103 which corresponds to the mean ellipticity of 0.31. Then, the additional contribution of external shear is estimated about 0.057 ~ 0.077, which can be thought to be originated from the other neighboring galaxies and dark matter. Therefore, based on our *SIS + shear* modeling process, we therefore conclude again that the internal ellipticity is a more dominant factor than the external shear for galaxy lensing from the *JVAS/CLASS* radio lens surveys.

Based on the Monte-Carlo calculations with the direct inverse mapping formula (*see* Section 2), we conclude that the main reason for the well-known puzzle of the observed *high Quads-to-Doubles ratio* is caused by the difference in their effectiveness for the lensing cross-section, which depends on its image multiplicity, while its corresponding magnification bias effects also depend on the survey selection bias. Since the shear and the magnification bias including survey selection limits, strongly reduce *the effective lensing* cross-section of 2-image systems compared to 4-image systems, our numerical analysis thus provides a straightforward insight on the core reason for *the anomaly of a high Quads-to-Doubles ratio* based on the simple, but representative, galaxy-lens model, the *SIS + shear* model. Thus, it would be worthwhile to state here that the shear term is a valuable parameter in modeling for both the point-lens (e.g. *Chang-Refsdal* lens) and extended galaxy-lens (e.g. *SIS + shear* lens) to conceptually explain their imaging properties instead of using complex elliptical models. As the shear works almost exactly like the ellipticity in galaxy lensing calculations, both terms can be internally inter-changeable to finally estimate the result as we derived above.

We present the illustration of caustics and multiple images for the purpose of comparing of a low- and a high-shear lensing in Section 3. Depending on the external shear, its caustic changes and imaging properties with the external shear and source sizes were illustrated. Moreover, we firstly introduced the imaging properties of this lens system, even when the external shear exceeds the critical value of 1.0. The maximum number of multiple images is found to be 3 (*Figure 6* in Section 3). Since such a special lensing property with a high-shear is not generally expected from normal galaxy-lensing environments, this could be useful for strongly perturbed galaxy-lensing events in the future. For example, in the case of cosmic-string lensing, Gott (1985) predicted that 2-image lensing having a comparable magnification ratio occurs, which is similar to the imaging property of the high-shear cases (*see Figure 6* in Section 3). We also found that the parities of 2-image systems in high-shears can produce identical parities of images, which is contrary to the low-shear lensing cases.

A lensing galaxy located nearby a cosmic-string should experience a strong tidal perturbation to the main lensing potential. Thus, further observational efforts for finding the cosmic lensing events are of a great interest in the future (e.g. Sazhin et al. 2007). We should note that our *effective lensing* cross-section analysis for the *SIS + shear* model should be valid for any other elliptical (*SIE*) models as well, as their dependencies of caustic properties depending on their external (or internal) perturbation are almost similar (Keeton, Kochanek & Seljack 1997). Also, our inverse mapping formula (*see* Section 2) of the *SIS + shear* lens model can be directly applied to various N-body galaxy-lensing calculations in the future, instead of using a simple pure *SIS* without shear model.

*Acknowledgments:* This work was supported by *the World CLASS University* (*W.C.U*) program and by *the BK plus* program through the National Research Foundation of Republic of Korea funded by *the Ministry of Education, Science & Technology* (R31-10016). DWL would like to give special thanks to Prof. *D.-H. Lee*, Prof. *K. Chang,* and to the memory of Prof. *S. Refsdal*.